\documentstyle[multicol,aps,prl,floats]{revtex}

\begin{document}
\title{On how a joint interaction of two innocent partners (smooth advection \&
linear damping) produces a strong intermittency}
\author{M. Chertkov}
\address{Physics Department, Princeton University, Princeton, NJ 08544}
\date{\today}
\maketitle

\begin{abstract}
Forced advection of passive scalar by a smooth $d$-dimensional
incompressible velocity in the presence of a linear damping is studied.
Acting separately advection \& dumping do not lead to an essential
intermittency of the steady scalar statistics, while being mixed together
produce a very strong non-Gaussianity in the convective range: $q$-th
(positive) moment of the absolute value of scalar difference, $\left\langle
|\theta (t;{\bf r})-\theta (t;0)|^{q}\right\rangle $ is proportional to $%
r^{\xi _{q}}$, $\xi _{q}=\sqrt{d^{2}/4+\alpha dq/\left[ (d-1)D\right] }-d/2$%
, where $\alpha /D$ measures the rate of the damping in the units of the
stretching rate. Probability density function (PDF) of the scalar difference
is also found.
\end{abstract}

\draft


Advection of passive scalar $\theta (t;{\bf r})$ by incompressible velocity
field is a classical problem in turbulence theory. The problem attracts a
lot of recent attention for remarkable combination of a rich physics and
nontrivial rigorous results derived. R.H. Kraichnan pioneered the rigorous
study of the problem inventing the temporal short-correlated but spatially
non-smooth model of velocity for which the simultaneous pair correlation
function of the scalar was found \cite{68Kra}. However, the question of
possible anomalous behavior of higher order ($n>1$) structure functions $%
S_{2n}(r)=\left\langle \left( \theta (t;r)-\theta (t;0)\right)
^{2n}\right\rangle \sim r^{\xi _{2n}}$ was posed only 25 years later \cite
{94Kra}. Next, the anomalous scaling $\Delta _{2n}\equiv n\xi _{2}-\xi _{2n}$%
, describing the law of the algebraic growth with $L/r$ (where $L$ is the
scale of the scalar pumping) of the dimensionless ratio $S_{2n}(r)/\left[
S_{2}(r)\right] ^{n}$, was shown to exist generically \cite
{95CFKLb,95GK,95SS}. The anomalous exponent was calculated perturbatively in
expansions about three non-anomalous ( $\Delta _{2n}=0$) limits, of large
space dimensionality $d$ \cite{95CFKLb,96CF}, of extremely non-smooth \cite
{95GK,96BGK} and almost smooth \cite{95SS} velocities respectively. A strong
anomalous scaling (saturation of $\xi _{2n}$ to a constant) was found for
the Kraichnan model at the largest $n$ by a steepest descent formalism \cite
{97Che}. Although the restricted asymptotic information about anomalous
exponent in the model is available a future possibility to establish a whole
rigorous dependence of $\xi _{2n}$ on $n$, $d$ and degree of velocity
non-smoothness seems very unlikely (in a sense, the recent Lagrangian
numerics \cite{98FMV} on the $\zeta _{4}$ as a function of $\zeta _{2}$ at $%
d=3$ compensates the lack of rigorous information).

In the present letter I discuss yet another passive scalar model with
nontrivial anomalous behavior, $\zeta _{2n}<n\xi _{2}$, which is possible to
resolve explicitly for all the values of the governing parameters. The model
describes generalization of the smooth (Batchelor) limit of the Kraichnan
model on the case of a linear damping of the scalar. The pure Batchelor
model (no damping), studied detaily in \cite
{59Bat,69Coc,74Kra,94SS,95CFKLa,94CGK,95BCKL,97BGK}, shows non-anomalous
(logarithmic and almost Gaussian for the Gaussian form of the pumping)
behavior. The limit of a huge linear damping (neglect advection) is also
non-anomalous. However, as it will be shown below (see (\ref{S2n},\ref{PDF}%
)), the anomalous scaling does exist generically, and appears to be a
nontrivial function of $n$, $d$ and a parameter standing for the
damping-to-convection ratio. The model describes forced advection of a
scalar pollutant in the viscous-convective range (the viscous-to-diffusivity
ratio is supposed to be large) absorbed instantly and homogeneously, for
example via a chemical reaction with other species present in the flow.
Therefore, all the predictions of the letter may be checked experimentally
as well as numerically.

Consider advection of passive scalar $\theta (t;{\bf r})$ by a smooth
incompressible velocity field, ${\bf u}(t;{\bf r})=\hat{\sigma}(t){\bf r}$ ($%
\hat{\sigma}(t)$ is $d\times d$ traceless matrix) in a presence of a linear
damping and diffusion, 
\begin{equation}
\partial _{t}\theta +\sigma ^{\mu \nu }(t)r^{\mu }\nabla _{r}^{\nu }\theta
=\kappa \triangle _{r}\theta -\alpha \theta +\phi .  \label{theta}
\end{equation}
The scalar is forced by random field $\phi (t;{\bf r})$, which for a sake of
simplicity is considered to be Gaussian thus fixed completely by $%
\left\langle \phi (t_{1};{\bf r}_{1})\phi \left( t_{2};{\bf r}_{2}\right)
\right\rangle $ $=\chi (|{\bf r}_{1}-{\bf r}_{2}|)\delta \left(
t_{1}-t_{2}\right) $, where the function $\chi (r)$ decays fast enough if $r$
exceeds the integral scale $L$. Although we are sure that many principal
results of the paper can be generalized for the case of a finite temporal
correlations of the strain matrix $\hat{\sigma}$, only the short correlated
one allowing a simple derivation is considered here. The Gaussian statistics
of $\hat{\sigma}$ is fixed by the pair correlation function $\left\langle
\sigma ^{\eta \mu }(t)\sigma ^{\beta \nu }(t^{\prime })\right\rangle $ equal
to 
\begin{equation}
D\left[ \left( d+1\right) \delta ^{\mu \nu }\delta ^{\eta \beta }-\delta
^{\mu \eta }\delta ^{\nu \beta }-\delta ^{\mu \beta }\delta ^{\nu \eta
}\right] \delta (t\!-\!t^{\prime }).  \nonumber
\end{equation}

We start studying the pair correlation function of the scalar field: $%
F(r_{12})=\left\langle \theta (t;{\bf r}_{1})\theta (t;{\bf r}%
_{2})\right\rangle $. Averaging two replicas of (\ref{theta}) multiplied by
the respective scalar counterparts one gets 
\begin{equation}
\left[ -r^{1-d}\partial _{r}r^{d}\left( D(d-1)r+\frac{2\kappa }{r}\right)
\partial _{r}+2\alpha \right] F(r)=\chi (r).  \label{F}
\end{equation}
Consider the case of a step-like pumping function, when $\chi (r)=P=const$
at $r<L$, and zero otherwise. Introduce a forced solution of this equation, $%
F_{f}(r)=P\vartheta (L-r)/[2\alpha ]$. Two zero modes of the operator from
the lhs of (\ref{F}) should be added to $F_{f}(r)$ to respect continuity of $%
F(r)$ and its derivative at $r=L$. Zero mode added at the upper ( $r>L$)
interval should vanish at $r\rightarrow L$, while its counterpart added at $%
r<L$ should be finite at the origin ( $r\rightarrow 0$). If the dissipative
scale, $r_{d}=\sqrt{\kappa /\max \{D,\alpha \}}$ is small enough, one gets 
\begin{equation}
F(r)=\frac{P}{2\alpha }\left\{ 
\begin{array}{c}
1-\frac{\xi _{-}}{\xi _{-}-\xi _{+}}\left( \frac{r}{L}\right) ^{\xi _{+}},%
\hspace{0.2in}r_{d}\ll r<L, \\ 
\frac{\xi _{+}}{\xi _{+}-\xi _{-}}\left( \frac{r}{L}\right) ^{\xi _{-}},%
\hspace{0.2in}r>L\gg r_{d};
\end{array}
\right.  \label{zetaPM}
\end{equation}
where $\xi _{\pm }\equiv \pm \sqrt{d^{2}/4+2\alpha d/\left[ \left(
d-1\right) D\right] }-d/2$. Finite diffusivity generalization of (\ref
{zetaPM}) (of its zero-mode part) can be presented in terms of the
hyper-geometric function. Dominant contribution into $S_{2}(r)$ at $r\ll L$
stems from a zero mode of the operator on the rhs of (\ref{F}) scaling as $%
r^{\xi _{+}}$.

Come to the study of the scalar difference stationary PDF, ${\cal P}\equiv
\left\langle \delta \left( x-\delta \theta _{r}\right) \right\rangle $.
Generally, at zero diffusion ($\kappa \rightarrow 0$) the stationary limit
is perfectly achieved via the direct balance between the pumping $\phi $ and
the $\alpha $-damping. Therefore, there is no dissipative anomaly in the
case and we may simply write the Fokker-Planck equation 
\begin{equation}
D(d-1)r^{1-d}\partial _{r}r^{d+1}\partial _{r}{\cal P}+\alpha \partial
_{x}\left( x{\cal P}\right) =0.  \label{FP}
\end{equation}
(\ref{FP}) is valid at $x<\theta _{L}$, where $\theta _{L}$ is the amplitude
of the scalar field at the integral scale, which is estimated by $P/\max
\{\alpha ,D\}$. Even without calculation of the PDF itself we may simply get
the anomalous exponents for the structure functions of all the orders, by
means of integration of (\ref{FP}) against the respective moments of $x$.
The only thing left is to pick up a vanishing at $r\rightarrow 0$ solution
of the linear ordinary homogeneous differential equation (zero mode of the
eddy-diffusivity operator). It gives for the even moments (odd moments are
constrained to be zero due to pumping isotropy and Gaussianity) 
\begin{equation}
\frac{S_{2n}(r)}{\theta _{L}^{2n}}\sim \left( \frac{r}{L}\right) ^{\xi
_{2n}},\hspace{0.2in}\xi _{2n}=\sqrt{\frac{d^{2}}{4}+\frac{2\alpha dn}{(d-1)D%
}}-\frac{d}{2},  \label{S2n}
\end{equation}
where the dependence on $L$ and $\theta _{L}$ is restored just by
dimensionality. (\ref{S2n}) is compatible with (\ref{zetaPM}) and it holds
for any $d$ , $n$, and $\alpha $. Notice that at the fixed value of the
damping-to-advection ratio $\alpha /D$ and $d$ going to infinity $\xi _{2n}$
goes to zero.

Although the calculation of anomalous exponents was our main goal the PDF
itself may be also simply extracted out of (\ref{FP}) if additionally the
form of the PDF at the integral scale (which in its own term relates to
concrete form of the pumping) is fixed. For example, the Gaussian PDF at the
integral scale corresponds to 
\begin{eqnarray}
{\cal P}\left( x;r\right)  &=&\frac{\ln \left[ L/r\right] }{\pi x\sqrt{a}}%
\left( \frac{L}{r}\right) ^{d/2}Q\left( \frac{x}{\theta _{L}};\ln \left[
L/r\right] \right) ,  \label{PDF} \\
Q\left( y;z\right)  &\equiv &\int\limits_{0}^{\ln \left[ 1/y\right] }\exp
\left[ -\frac{z^{2}}{4at}-\frac{d^{2}a}{4}t-y^{2}\exp \left[ 2t\right]
\right] \frac{dt}{t^{3/2}},  \nonumber
\end{eqnarray}
where $a\equiv D(d-1)/\alpha $. At $L\gg r$ the PDF shows a change in
behavior about $x_{c}\equiv \theta _{L}\left[ r/L\right] ^{1/(da)}$. $%
Q\left( y;z\right) $ is finite at the origin, $Q\left( 0;z\right) \sim
\left( r/L\right) ^{d/2}/\ln \left[ L/r\right] $, and further the alrgebraic
in $x$ decay at $x\ll x_{c}$, $Q\left( y;z\right) -Q\left( 0;z\right) \sim
y^{d^{2}a/4}$, turns into $Q\left( y;z\right) \sim y^{d^{2}a/4}\exp \left[
-z^{2}/\left( a\ln \left[ 1/y\right] \right) \right] $ at $x\gg x_{c}$.
Therefore, (\ref{S2n}) is applicable for all the positive (not necessarily
integer) moments of $|\delta \theta _{r}|$. All the negative moments are
divergent.

The possibility of two-point consideration explained above is based on the
absence of the dissipative anomaly. To prove this and also to shed some
light on the dynamical origin of anomalous behavior we consider Lagrangian
multi-point representation of the problem and show how does it lead to the
same answer (\ref{S2n}) for the asymptotic behavior of the structure
functions. (\ref{theta}) is equivalent to 
\begin{equation}
\left\{ 
\begin{array}{c}
\theta (t;{\bf r})=\int\limits_{0}^{\infty }dt^{\prime }\exp \left[ -\alpha
t^{\prime }\right] \phi \left( t^{\prime };{\bf \rho }(t-t^{\prime })\right)
\\ 
\frac{d}{dt}{\bf \rho }(t)=\hat{\sigma}(t){\bf \rho }(t)+{\bf \eta }(t),%
\hspace{0.2in}{\bf \rho }(0)={\bf r,}
\end{array}
\right.  \label{Lagr}
\end{equation}
where ${\bf \eta }(t)$ is the Langevin noise fixed by, $\left\langle \eta
^{\alpha }(t)\eta ^{\beta }(t^{\prime })\right\rangle =2\kappa \delta
(t-t^{\prime })$. Averaging the simultaneous product of $2n$ different
replicas of (\ref{Lagr}) one gets $F_{1\cdots 2n}\equiv \left\langle \theta
_{1}\cdots \theta _{2n}\right\rangle $. Once the average of the multi point
product is known it is easy to construct the desirable structure function $%
S_{2n}(r)$ fusing the $2n$ points. Moreover, one can get $S_{2n}$ from the
general object with all the points ${\bf r}_{i}$ being placed along a
straight line. Also we expect, and it will be confirmed below, that at $%
\alpha >0$ all the fused averages are finite also in the limit of zero
diffusivity. The last observation allows to consider the infinite Peclet
number $Pe\equiv L/r_{d}$ limit just replacing $\kappa $ by zero. Therefore,
for the collinear geometry ${\bf r}_{i}={\bf n}$ $r_{i}$ and at $%
Pe\rightarrow \infty $ direct averaging of (\ref{Lagr}) with respect to
statistics of $\phi $ and $\hat{\sigma}$ gives 
\begin{equation}
F_{1\cdots 2n}=\sum\limits_{\{i_{1,\cdots },i_{2n}\}}^{\{1,\cdots
,2n\}}\left\langle \prod\limits_{k=1}^{n}\int\limits_{0}^{\infty
}dt_{k}e^{-\alpha t_{k}}\chi \left[ e^{\eta (t_{k})}r_{i_{k};i_{k+1}}\right]
\right\rangle _{\eta },  \label{theta3}
\end{equation}
where $\eta (t)\equiv |\hat{W}(t){\bf n}|$ and $\hat{W}(t)$ satisfies $d\hat{%
W}(t)/dt=\hat{\sigma}(t)\hat{W}(t)$. The $\alpha =0$ version of (\ref{theta3}%
) was calculated in \cite{95CFKLa} for the $d=2$ case and generalized for
any $d\geq 2$ in \cite{94CGK} via a change of variables and further
straightforward transformation of the path integral standing for the average
over $\hat{\sigma}(t)$. Average with respect to all the ''angular'' degree
of freedoms gives the following effective measure of averaging with respect
to the rate of stretching along the ${\bf n}$ direction \cite{95CFKLa,94CGK} 
\[
{\cal D}\eta (t)\exp \left[ -d\int\limits_{0}^{\infty }dt\frac{\dot{\eta}%
^{2}+D^{2}(d-1)^{2}-2(d-1)\dot{\eta}D}{4(d-1)D}\right] . 
\]
Therefore, averaging with respect to the fluctuations of the $\eta $ field
produces 
\begin{eqnarray}
F_{1\cdots 2n} &=&n!\int\limits_{0}^{\infty
}dt_{1}\int\limits_{0}^{t_{1}}dt_{2}\cdots
\int\limits_{0}^{t_{n-1}}dt_{n}\int_{-\infty }^{+\infty }d\eta _{1}\cdots
d\eta _{n}\exp \left[ \frac{d}{2}\eta _{1}-\frac{D(d-1)d}{4}t_{1}\right] 
\nonumber \\
&&\times \sum\limits_{\{k_{i},\cdots ,k_{2n}\}}^{\{1,\cdots
,2n\}}\prod_{i=1}^{n}\left[ e^{2\alpha t_{i}}\chi \left( e^{\eta
_{i}}r_{k_{2i},k_{2i+1}}\right) G\left( t_{i-1}-t_{i};\eta _{i-1}-\eta
_{i}\right) \right] ,  \label{expr}
\end{eqnarray}
where $t_{i+1}=\eta _{i+1}=0$ and $G\left( t;\eta \right) \equiv \sqrt{\frac{%
\pi d}{4(d-1)tD}}\exp \left[ -\frac{d\eta ^{2}}{4(d-1)tD}\right] $ is the
Green function of the diffusive kernel. The integrand of (\ref{expr}) decays
exponentially in time, so the major contribution into the object forms at $%
t_{i}\sim 1/\alpha $, it does not depend on any $r_{ij}$. The first $r$%
-dependent contribution stems from $n-1$ temporal integrals formed at $\tau
\sim 1/\alpha $, and one at $t_{i}\sim \tau _{r}\sim \ln \left[ L/r\right]
/\max \{\alpha ,D\}$ (this special integration brings a spatial dependence
into the object, therefore on a single distance). Generally, there exists a
variety of terms with all the possible combinations, like a term with $k$
integration formed at $\tau $, while $n-k$ ones at $\tau _{r}$, and
therefore dependent explicitly on $2(n-k)$ points. However, we are looking
exclusively for a term dependent on all the $2n$ points since only such a
term of (\ref{expr}) contributes $S_{2n}(r)$. And it is really simple to
calculate the scaling of the term making use of the temporal scale
separation, $\tau _{r}\gg \tau $. Indeed, the large time contribution may be
extracted out of (\ref{expr}) in a saddle-point calculation. Variation of
all the exponential terms in (\ref{expr}) with respect to $t_{i}$ gives a
chain of saddle equations. The $\chi $ functions in the integrand of (\ref
{expr}) limits the $\eta $ integrations from above by $\ln \left[ L/r\right]
.$ Therefore, the desirable $2n$-points contribution forms at $t_{i}=\sqrt{%
d/\left[ 4(d-1)D\left( 2\alpha n+D(d-1)d/4\right) \right] }\ln \left[
L/r\right] $, and $\eta _{i}=\ln \left[ L/r\right] $, where in the leading
logarithmic order one should not distinguish between contributions of
different separations $r_{ij}$. Finally, substituting the saddle-point
values of $t_{i}$ and $\eta _{i}$ into (\ref{expr}) we arrive at (\ref{S2n}).

The basic physics of nonzero $\xi _{2n}$ (means deviating from the naive
balance of pumping and advection) and generally anomalous ($\Delta _{2n}\neq
0$) scaling at $\alpha >0$ can be stated quite directly. According to (\ref
{theta3}) the advection changes scales but not amplitude, while the
amplitude of injected scalar field decays exponentially from the time of
injection at the constant rate $\alpha $. The temporal integrals in (\ref
{expr}) forms at the mean time to reach a scale which is proportional to the
negative log of the scale. However, the effective spread in the factor by
which amplitude has decayed, upon reaching a given scale, increases as scale
decreases. It is why $\xi _{2n}>0$. Also there is more room for fluctuations
about the mean time due to the interference between the exponential decay of
the scalar amplitude and fluctuations of the stretching rate $\eta $. Thus
intermittency increases with decrease of scale size.

I conclude by couple of general remarks. First of all, the model gives an
example of the situation when the dissipative anomaly is absent, while the
intermittency (anomalous scaling, $\Delta _{2n}\neq 0$) takes place. Second,
continuous dependance of the exponents on the damping rate originates from
coincidence of the scaling dimensions (zero in the Batchelor case) of the
bare eddy diffusivity operator and the damping-dependent correction to it.

I thank I. Kolokolov, R. Pierrehumbert, B. Shraiman for inspiring
discussions. Valuable comments of G. Falkovich, R. Kraichnan, and M.
Vergassola are greatly appreciated. The work was supported by a R.H. Dicke
fellowship.


\end{document}